\documentclass[12pt]{article}
\usepackage{epsfig}

\newcommand{\be}{\begin{equation}}
\newcommand{\ee}{\end{equation}}
\newcommand{\beqn}{\begin{eqnarray}}
\newcommand{\eeqn}{\end{eqnarray}}

\newcommand{\Slash}[1]{\ooalign{\hfil/\hfil \crcr$#1$}}
\newcommand{\itep}{~\vspace{-1.5cm}

\begin{flushright}
{\large HU-EP-02-62}\\
{\large ITEP-LAT/2003-02}\\
{\large January 13, 2003}
\end{flushright}
\vspace{1.0cm}}

\begin{document}

\thispagestyle{empty}

\baselineskip=14pt

\begin{center}

\itep

{\Large\bf
Evidence for Dyonic Structure \\
\vspace{2mm}
of $SU(2)$ Lattice Gauge Fields Below $T_c$   
\footnote{Work supported by DFG. 
B.~V.~M. and A.~I.~V. acknowledge receiving grants 
RFBR 02-02-17308 and 01-02-17456, INTAS 00-00111 and
the CRDF award RP1-2364-MO-02. } 
}

\vskip 1.0cm
{\large
E.--M.~Ilgenfritz$^a$\footnote{Talk presented by E.-M.~I. 
at the 5th International Conference on Quark Confinement and the Hadron Spectrum, 
Gargnano, Italy, 10-14 Sep 2002}, 
B.~V.~Martemyanov$^b$, M.~M\"uller--Preussker$^a$, 
S.~Shcheredin$^a$ and A.~I.~Veselov$^b$
}\\

\vspace{.4cm}

{ \it
$^a$ Institut f\"ur Physik, Humboldt--Universit\"at zu Berlin,
D-10115 Berlin, Germany

\vspace{0.3cm}

$^b$ ITEP, B. Cheremushkinskaya 25, Moscow, 117259, Russia

}
\end{center}

\begin{abstract}
{We report the observation of calorons with nontrivial
holonomy and fractional topological charge objects in cooled  
lattice samples derived from $SU(2)$ equilibrium ensembles at $T<T_c$. 
}
\end{abstract}

Characteristic features of QCD are related to the existence of lumps of 
topological charge in the vacuum. Usually, they are described as instantons
or, at $T \ne 0$, calorons~\cite{GPY} and considered to be the agents 
of chiral symmetry breaking. 
Recently, this view has been questioned from different sides.
At first, an old companion of instantons, the meron picture (see {\it 
e.g.}~\cite{meron}) has been revived. One motivation is to deduce confinement 
from the dissociation of instantons into constituents. The semiclassical 
picture of instantons has been totally challenged by Horvath et al.~\cite{Horvath}. 
Van Baal et al.~\cite{KvB} have stressed the existence of more general 
instanton solutions at $T \ne 0$, distinguished by nontrivial holonomy boundary 
conditions. A semiclassical approach to deal with this situation has been outlined 
for four dimensional ${\cal N} = 1$ supersymmetric gluodynamics~\cite{Mattis}. 

The present talk reports on recent lattice evidence~\cite{IMMSV} for the existence 
of dyonic lumps in the confinement phase. They are expected to be constituents 
of the new caloron solutions and show the required color correlations.
The new (KvB) caloron~\cite{KvB} is a finite-temperature (periodic) solution 
of the Euclidean field equations. It is selfdual with topological charge 
$Q=\pm 1$. The holonomy at asymptotics is not restricted to trivial values, 
$P=\exp \left( i \int_0^{1/T} A_4 dx_4 \right) \to P_{\infty} \Slash{\in} Z(N_c)$. 
Depending on $P_{\infty}$, part of the parameter space corresponds to 
solutions with action (and topological charge) concentrated in $N_c$ 
well-separated static constituents. One of the constituents is carrying 
the fermion zero eigenmode required to exist for a $Q=\pm 1$ object 
according to the Atiyah-Singer index theorem.

\begin{figure}[!htb]
\vspace{3mm}
\begin{center}
\begin{tabular}{cc}
\epsfxsize=8.0cm \epsffile{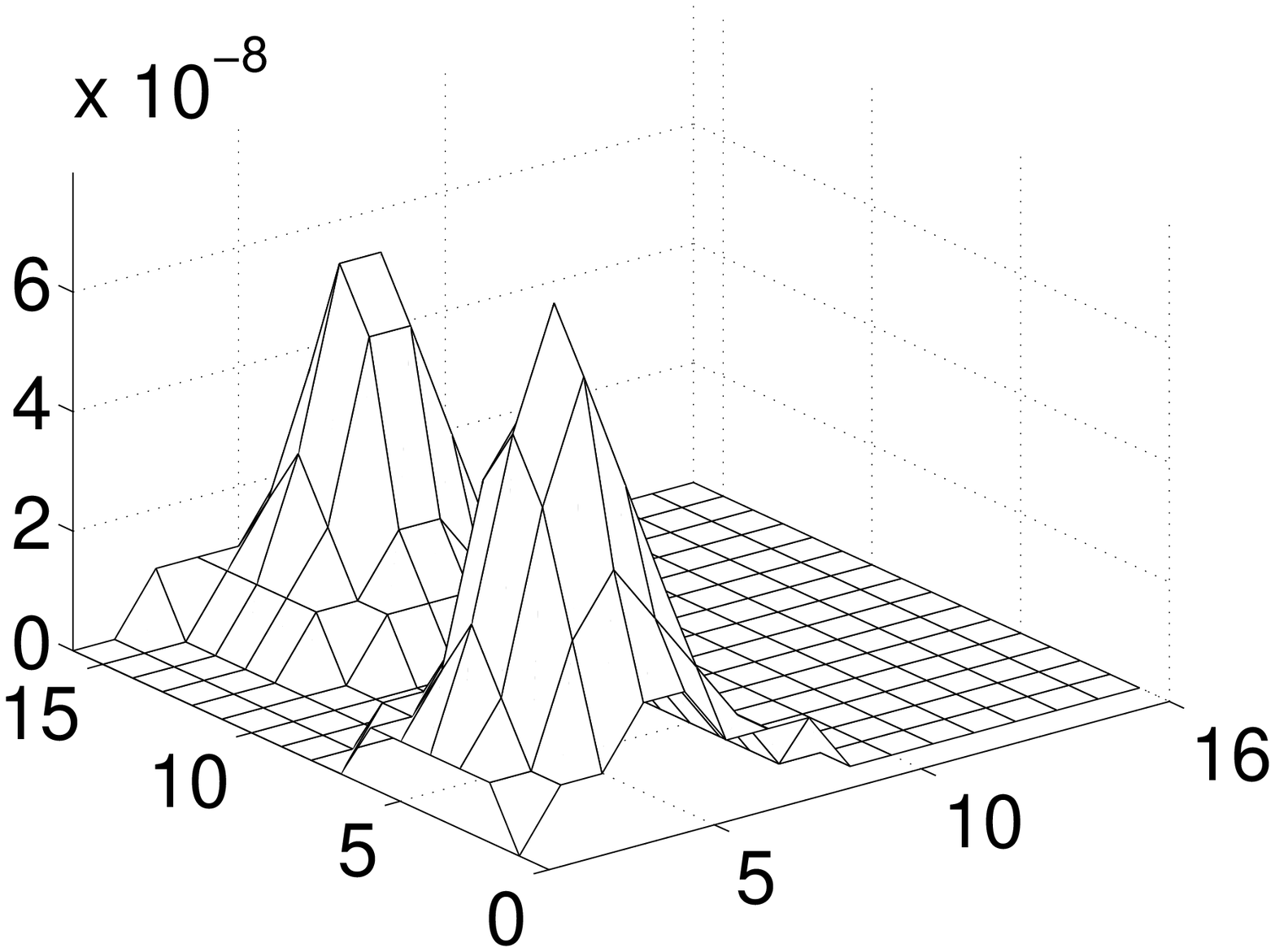} &
\epsfxsize=8.0cm \epsffile{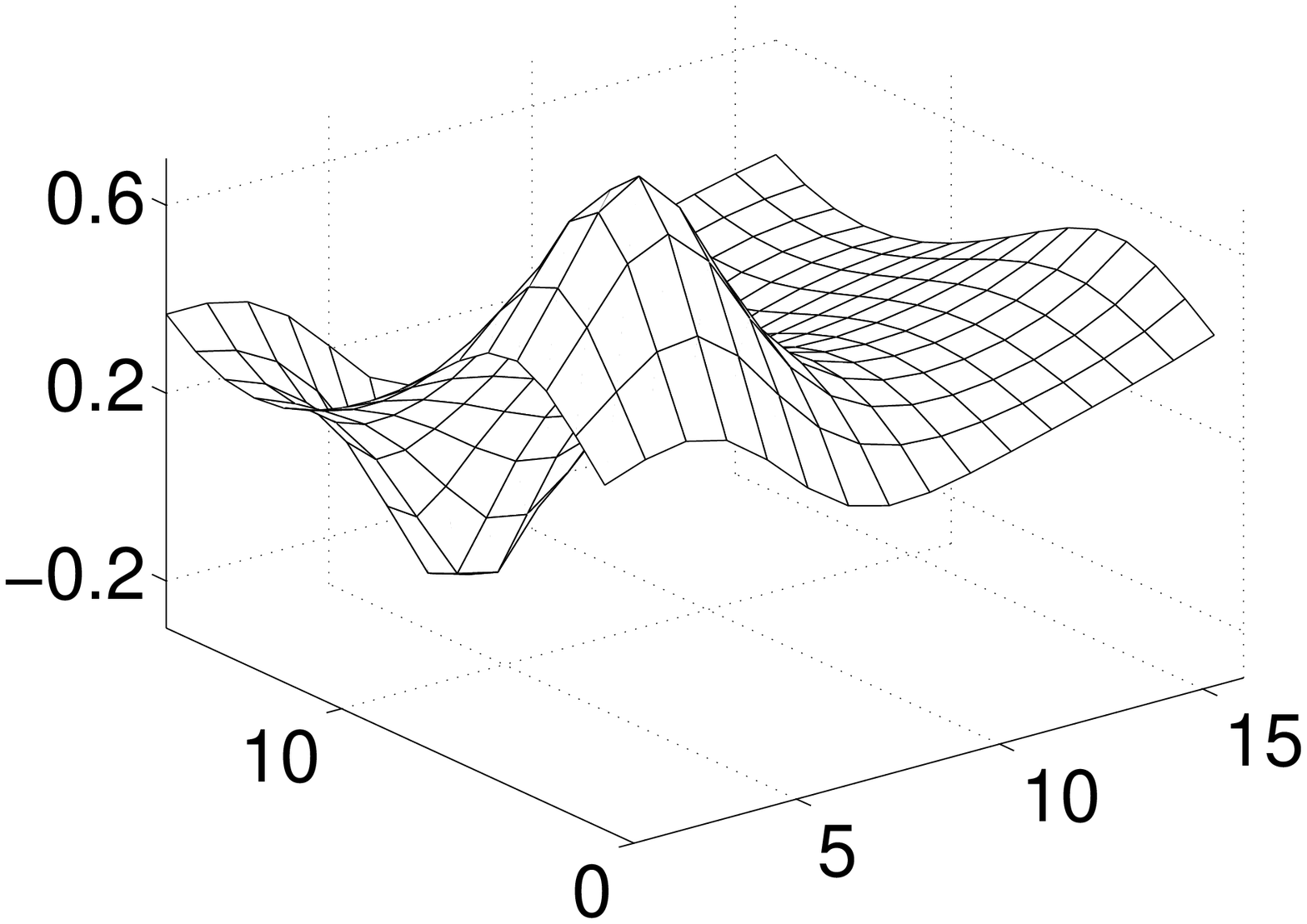} \\
\end{tabular}
\end{center}
\caption{Topological density $q(x)$ (left) 
         and Polyakov loop $L({\vec x})$ (right)
	 of a dyon-dyon pair
\label{profile}}
\end{figure}

\begin{figure}[!htb]
\vspace{3mm}
\begin{center}
\begin{tabular}{cc}
\epsfxsize=8.0cm \epsffile{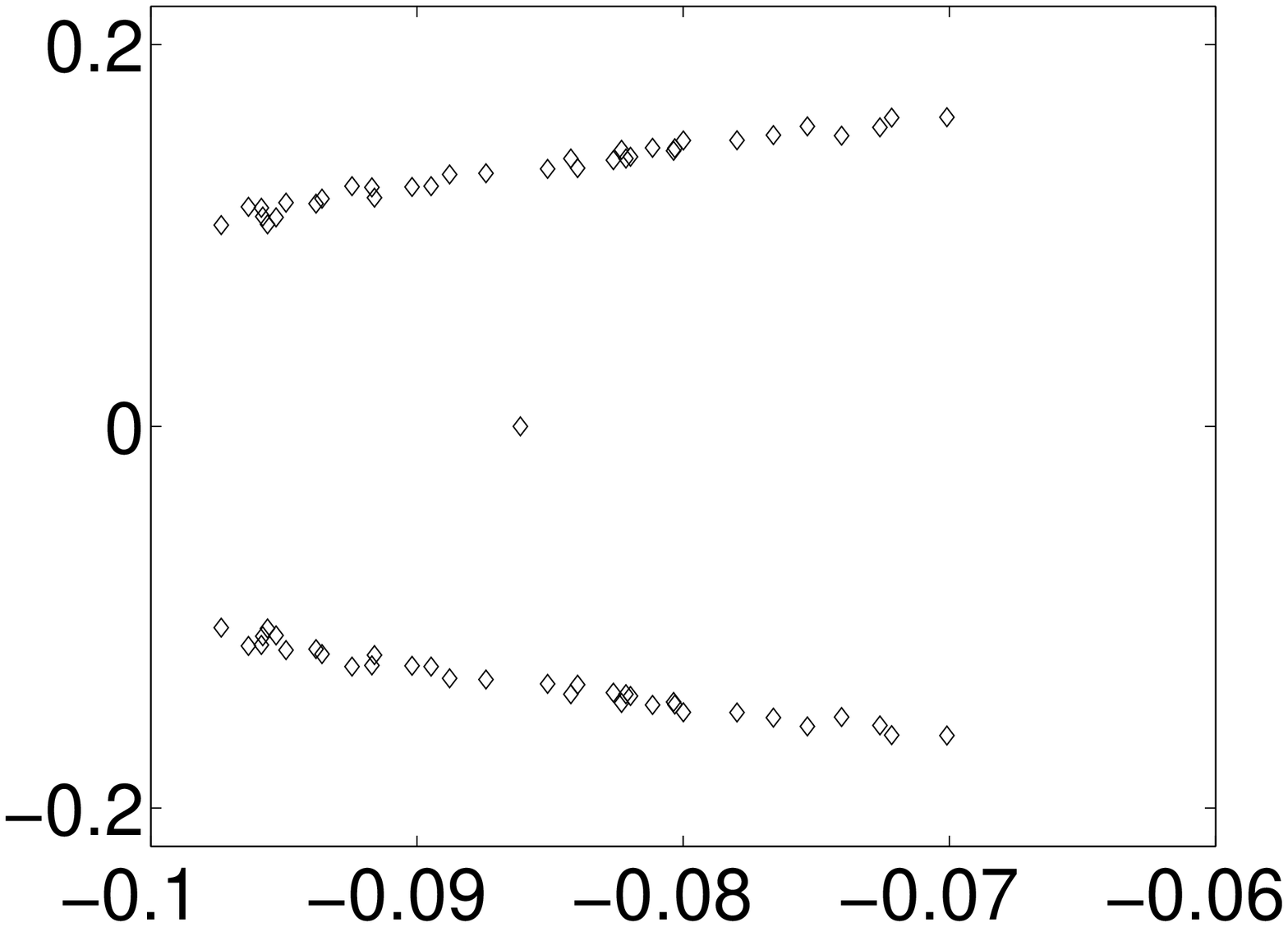} &
\epsfxsize=8.0cm \epsffile{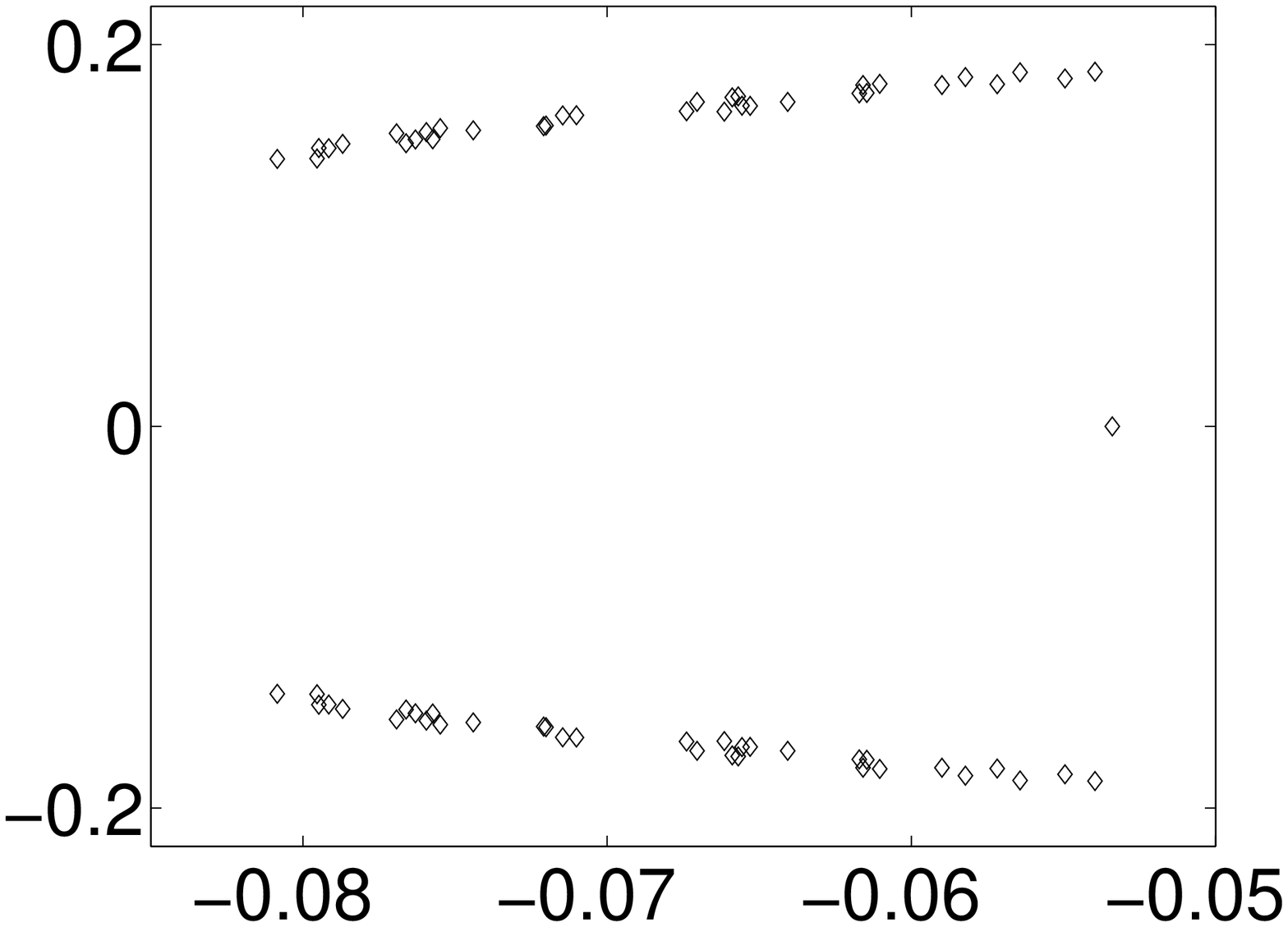} \\
\end{tabular}
\end{center}
\caption{Lowest fermionic eigenvalues for periodic (left)
         and antiperiodic (right) boundary conditions
	 in Euclidean time
\label{eigenvalue}}
\end{figure}

\begin{figure}[!htb]
\vspace{6mm}
\begin{center}
\begin{tabular}{cc}
\epsfxsize=8.0cm \epsffile{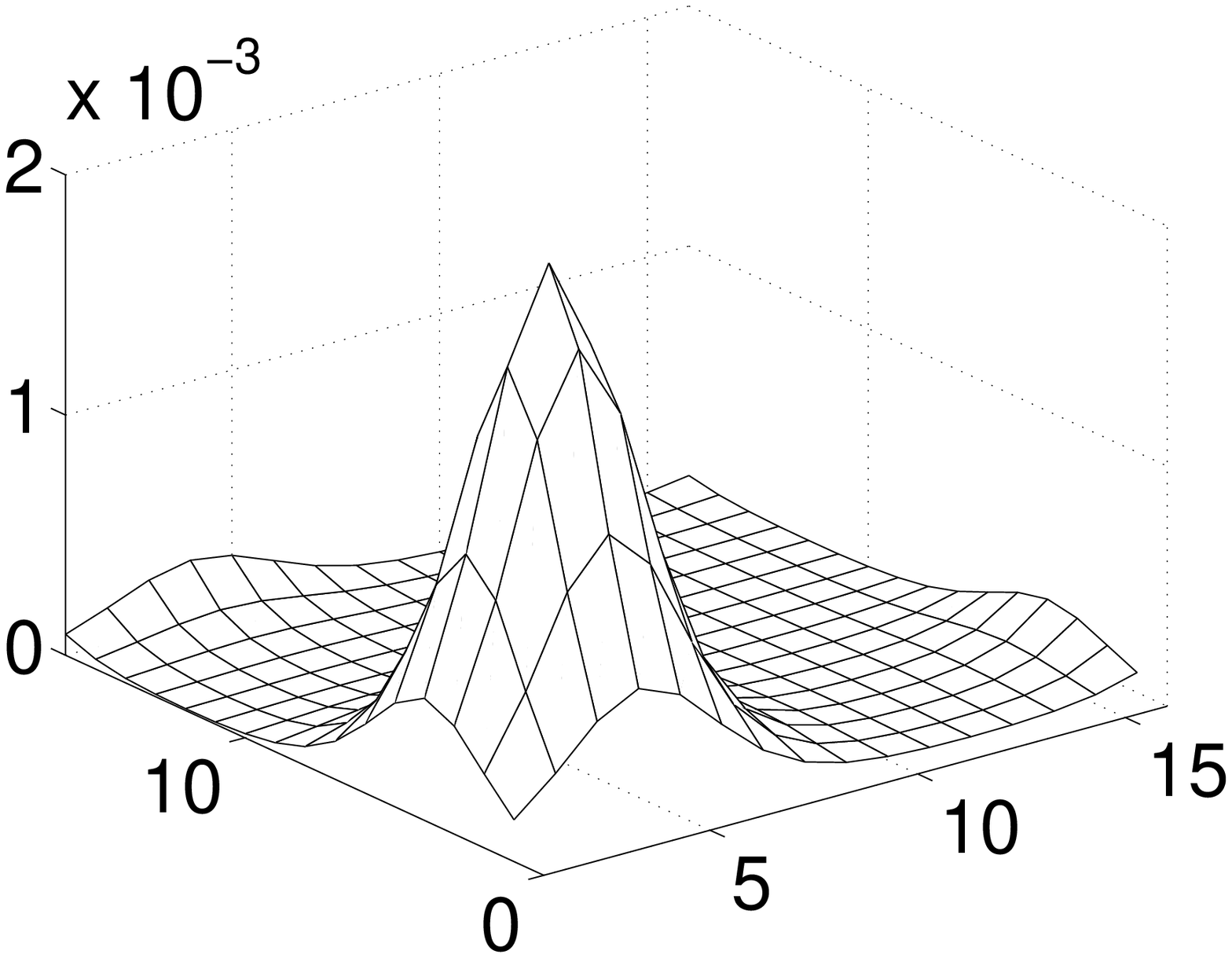} &
\epsfxsize=8.0cm \epsffile{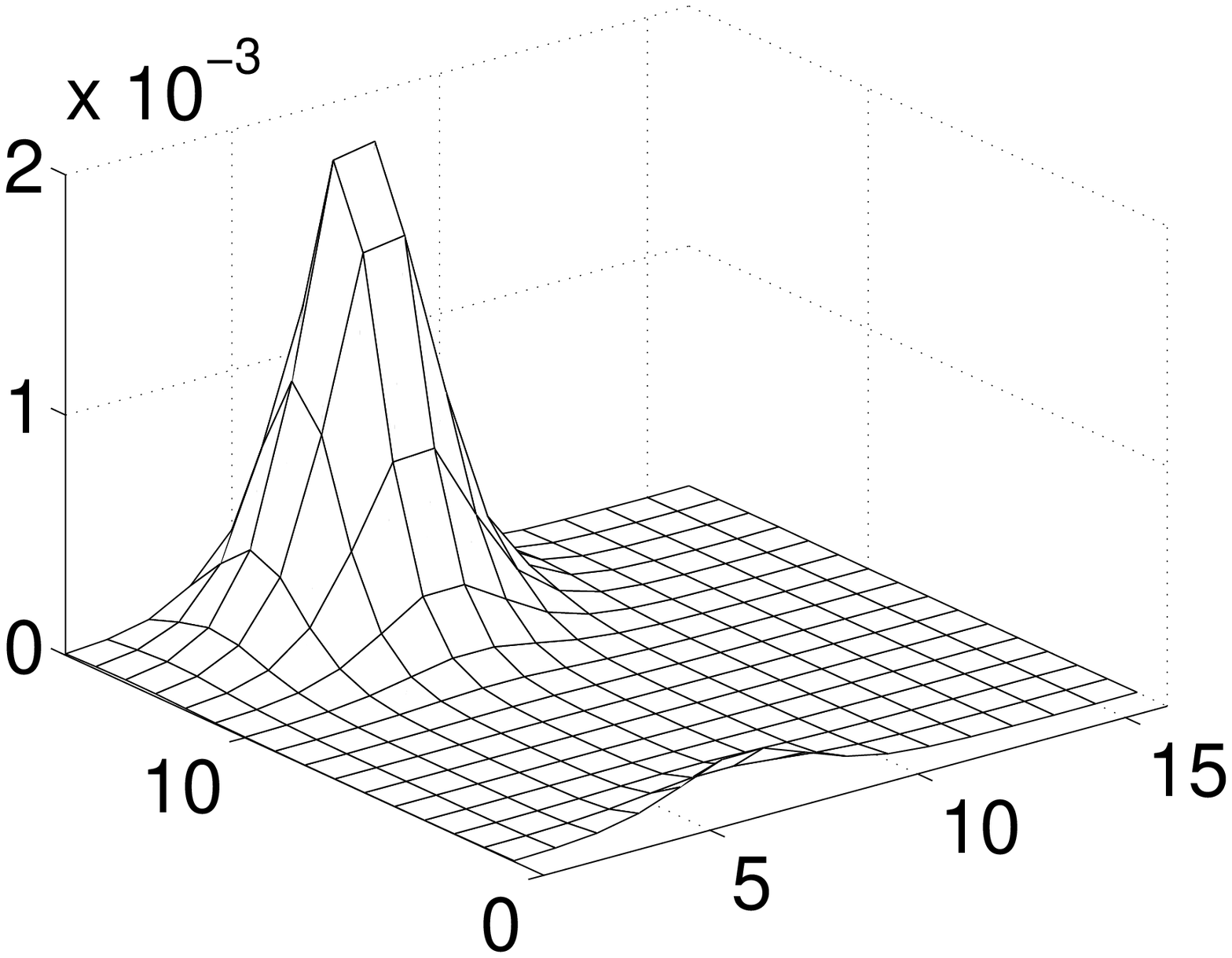} \\
\end{tabular}
\end{center}
\caption{Localized fermionic zero mode, for periodic 
         (left) and antiperiodic (right) boundary conditions
	 in Euclidean time
\label{localized}}
\end{figure}

Monte Carlo configurations taken from confined and unconfined ensembles, 
respectively, have been ''cooled'' (by gradual minimization 
of action). We wanted to show~\cite{IMMSV} that nearly classical 
configurations of KvB type really pop up as background fields starting 
from the confinement phase. In fact, this happens at temperature close 
to $T_c$. In contrast, the deconfined phase does not support topologically 
nontrivial background fields. Monitoring the relaxation 
$\frac{dA}{d\tau} \sim -\frac{\partial S}{\partial A}$, cooling was stopped 
triggered by the gradient becoming {\it minimal within some window of action},
and snapshots of configurations were taken. On plateaux at moderate 
multiples of $S_{inst}=2\pi^2$, the multi-lump structure is still comprehensible 
and can be analysed as described in the following.   
Figs. \ref{profile} - \ref{localized} show different portraits of 
the same selfdual {\it dyon-dyon pair} obtained by cooling under periodic gluonic 
boundary conditions with $S \approx 2\pi^2$. 
Almost static dyons are localizable by peaks of the topological 
density $q(x)>0$ and the Polyakov loop $L({\vec x}) \approx \pm 1$. 

A complementary view is given by the spectrum of the (non-Hermitean) Wilson 
fermion Dirac operator. Real eigenvalues refer to a fermionic zero mode 
which, depending on periodic (p.b.c.) or antiperiodic (a.b.c.) fermion 
boundary conditions in time, is localized at one or the other constituent. 
Cooled configurations like this can be well fitted by the shape functions of 
analytic KvB solutions. Nonstatic $Q=1$ caloron configurations can also 
be classified according to the KvB solutions. 
Unexpectedly, also {\it dyon-antidyon pairs} (with opposite topological charge 
but same sign Polyakov loop) have been detected in the same window of action and 
found to be almost stable. In this case a pair of almost-real modes exists, 
similar to those of an instanton-antiinstanton pair.
The saddle point configurations in higher windows of action (and their cascade 
decay under cooling) can be interpreted in terms of such dyonic constituents. 

Further details can be found in Ref.\cite{IMMSV}. 
Corroborating the point of view dealt with here, {\it fermionic} results 
concerning {\it uncooled} configurations of $SU(3)$ gluodynamics have been 
reported~\cite{Gattringer1,Gattringer2} after the conference. 

\vspace{-2mm}

\begin{thebibliography}{99} 

\bibitem{GPY}
D.~J.~Gross, R.~D.~Pisarski, L.~G.~Yaffe, Rev.~Mod.~Phys.~{\bf 53} (1981) 43. 

\bibitem{meron}
A.~Montero and J.~W.~Negele, Phys.~Lett.~{\bf B 533} (2002) 322.

\bibitem{Horvath}
I.~Horvath et al., these proceedings and references therein.

\bibitem{KvB}
T.~C.~Kraan and P.~van~Baal, Phys.~Lett.~{\bf B 435} (1998) 389, 
Nucl.~Phys.~{\bf B 533} (1998) 627.

\bibitem{Mattis}
N.~M.~Davies et al., Nucl.~Phys.~{\bf B 559} (1999) 123.

\bibitem{IMMSV}
E.--M.~Ilgenfritz et al., Phys.~Rev.~{\bf D 66} (2002) 074503.

\bibitem{Gattringer1}
C. Gattringer, e-print hep-lat/0210001, Phys.~Rev.~{\bf D} in print.

\bibitem{Gattringer2}
C. Gattringer and S. Schaefer, e-print hep-lat/0212029.

\end{thebibliography}
\end{document}